\documentclass[a4paper]{amsart}             
\usepackage{amssymb}               
\usepackage{verbatim}             
\usepackage{graphicx}    
\usepackage{upref}    
\usepackage[latin1]{inputenc}  

\makeatletter             
\let\latex@makecaption\@makecaption                
\def\@makecaption{\small\latex@makecaption}             
\makeatother              
\allowdisplaybreaks                  
\newtheorem{theorem}{Theorem}[section]

\begin{document}                
 \title[ Multiple model-free knockoffs ]{Multiple model-free knockoffs }                 
% Information for  author                 
%\author[H.\ Holden]{Helge Holden}                
%\address{Department of Mathematical Sciences,                
%Norwegian University of                
%Science and Technology, NO--7491 Trondheim, Norway,  
%and \newline  
%  Centre of Mathematics for Applications,   
% % Department of Mathematics,  
%  University of Oslo,  
%  P.O.\ Box 1053, Blindern,  
%  NO--0316 Oslo, Norway}                  
%\email{holden@math.ntnu.no}                  
%\urladdr{http://www.math.ntnu.no/\~{}holden/}   
\author[L.\ Holden, K.H. Hellton]{Lars Holden, Kristoffer H.\ Hellton }                   
\address{Norwegian Computing    
Center, P.\ O.\ Box 114 Blindern,                   
NO--0314 Oslo, Norway}                   
\email{lars.holden@nr.no}                   
 \date{\today\ } % ({\it Preliminary version})}      
%\keywords{Diversification, Value at Risk, Business area}   
%\subjclass[2000]{}                
\begin{abstract}  
Model-free knockoffs is a recently proposed technique for identifying covariates that is likely
to have an effect on a response variable. The method is an efficient method to control the false
discovery rate in hypothesis tests for separate covariates. 
 This paper presents a generalization  of the technique  using multiple sets of model-free  knockoffs. 
This is formulated as an open question in  Candes et al.  \cite{Candes17}. 
With multiple knockoffs, we are able to reduce the randomness in the knockoffs, 
making the result stronger. Since we use the same structure for generating all the knockoffs, 
the computational resources is far smaller than proportional with the number of knockoffs. 

%  {\bf We need some more realistic examples showing the improvement with several set of knockoffs.} 

\end{abstract}                   
                   
\maketitle

%\section*{Status}                  
%\begin{enumerate}                      
%\item New section 6.                
%\end{enumerate}                       
                  
%------------------- section                  
\section{Introduction}       \label{intro}                
   
Many applications have a large number of potential covariates that may influence the response
 variable of interest. The standard method to reduce the number of covariates, is to perform hypothesis
tests for each of the covariates in order to identify a small number of covariates where we can reject
the hypothesis that the covariate have no influence on the response variable. The challenge is to control 
the false discovery rate, FDR, in these hypothesis tests. 
In a series of recent papers \cite{Barber15}, \cite{Candes17}, \cite{Barber17}
  and  \cite{Barber18} a new method denoted model-free knockoffs is presented. This approach assumes that
we known the joint distribution of the covariates, but makes no assumptions on the relationship between
the covariates and the response function. Further, we assume that the observations of the response are 
independent and identically distributed.

We develop this new approach further by applying several sets of independent and identically distributed
sets of knockoffs. This was proposed as an open question in Candes et al.  \cite{Candes17}. This approach
increases the power of the method by reducing the randomness in the simulated knockoffs. We introduce a new
assumption and based on this assumption, we prove
stronger bounds for FDR when the number of sets of knockoffs increases. 
Except for proposing several sets of knockoffs, we
follow the approach in Candes et al.  \cite{Candes17} closely.          

We assume a typical linear regression model $$y=X \beta + \varepsilon$$ where
$y \in R^{n}$ is a vector of response, $X \in R^{n \times p}$ is a known design matrix, $\beta \in R^{p}$ is an 
unknown vector of coefficients and $\varepsilon \sim N(0, \sigma^{2} I )$  is Gaussian noise. 
Both $n$ and $p$ may be large and there is limited data, i.e.  we don't expect
 $n \gg p$. We  may have $n<p$.

It is natural as a hypothesis, to assume that each covariate does not influence the response variable. 
Based on data, we will reject this hypothesis for some of the covariates. 
More formally,  we define the set $S$ of covariates where $j \not\in S$ if the response $Y$ is independent
of  $X_j$ conditionally  on all the other variables $X_{-j}$. $S$ is denoted the Markov blanket of $Y$, see
 \cite{Candes17}. Our hypothesis is that $j \in H_0 = \Omega \backslash S $ where $\Omega$ is the set of 
all covariates. Our  estimate $\hat{S}$ of $S$ consists of the covariates where we 
reject the hypothesis that $j \in H_0$. We want to control the FDR by
 $E \#(j: j \in (\hat{S}\backslash S))/\# (j: j \in \hat{S}) \leq q$ for a given constant $q<1$.   
%Hence, our ambition is to identify a limited number
%of $\beta_j$ where we reject the hypothesis that $\beta_j=0$.  
%We want to ensure that the ratio of the covariates  where we 
%claim that $j \notin H_0 $,  where in fact $j \in H_0 $ is bounded by $q$.

 \section{Model} 
We follow the approach of Barber and Candes \cite{Barber15} and Candes et al.  \cite{Candes17} but instead
of  making one set of knockoffs, 
 we will generate $k$ sets of independent
 knockoffs. Each set of knockoffs has the same property as in Candes et al.  \cite{Candes17}.
In  \cite{Candes17} it is assumed a perfect knowledge of the distribution of $X$. This is generalized to 
an approximate knowledge of $X$ in  \cite{Barber18}. We follow the approach in \cite{Candes17} but our
approach works equally well for the assumptions in  \cite{Barber18}. 

 Under a Gaussian assumption the 
sets of knockoffs may be performed as follows: 

Assume that the original covariates have the form $X \sim N(0, \Sigma)$ after a normalization. The
covariates for each set of  knockoff $\tilde{X}_i$ for $i=1,2,\cdots , k$ must satisfy
$$ (X,\tilde{X}_i) \sim N(0,G) $$
where 
$$ G= \begin{bmatrix} \Sigma & \Sigma - diag\{ s \}  \\ \Sigma - diag\{ s \}  & \Sigma  \end{bmatrix} . $$
The knockoffs may be simulated from the distribution
\begin{equation}
\tilde{X}_i \mid X  \sim N(\mu,V) \label{knockoff}
\end{equation}

where $\mu$ and $V$ satisfy the standard regression formulas: 
$$\mu = X -X \Sigma^{-1} diag\{s\}  $$
$$V = 2 diag\{ s \} X - diag\{ s \} \Sigma^{-1} diag\{ s \} . $$
Here $ diag\{ s \}$ is any diagonal matrix such that $V$ is positive definite. However, we know that the 
strength of our prediction increases when the elements in $diag\{s  \}$ increase since this reduces the
dependence between $X_j$ and $\tilde{X}_{j,i}$ for each component $j$.

Based on the covariates $(X,\tilde{X}_i)$, we may generate the variables $Z_i=(Z_{1,i}, \cdots ,Z_{p,i} )$ and the knockoff variables
 $\tilde{Z}_i = (\tilde{Z}_{1,i}, \cdots, \tilde{Z}_{p,i})$. These are typically generated from t-statistics as
$$T_i =(Z_{1,i}, \cdots ,Z_{p,i}, \tilde{Z}_{1,i}\cdots, \tilde{Z}_{p,i} ) =t((X,\tilde{X}_i),y) $$
where $y$ is the response variable in the data and $i=1, \cdots , k$. 

 The t-statistics is typically the absolute value of the 
estimated Lasso coefficient of the component.  Then a large absolute value of  $Z_{j,i}$
indicates that component $j$ is significant while we know that the value of $\tilde{Z}_{j,i}$
 is independent of whether there is signal in component $j$. 
Compared to the notation in Candes et al.  \cite{Candes17}   $Z_{j,i}$ corresponds to $Z_{j}$
and  $ \tilde{Z}_{j,i}$   
correspond to   $ \tilde{Z}_{j}$ for $i=1, \cdots , 2k-1$. 
All the knockoff variables  $ \tilde{Z}_{j,i}$  may be generated with the same
$s$ in Candes et al. making the computational resources necessary for generating the knockoffs far
smaller than proportional with the number of knockoffs.

Define the sets of knockoffs $\tilde{Z}_{j,i}$ for $i=1,2,\cdots, 2k-1$ and 
 the statistics 
\begin{equation}   
W_{j,1}=Z_{j,1}- \frac{1}{k-1} \sum_{u=k+1 }^{2k-1}\tilde{Z}_{j,u}
\end{equation}
and
\begin{equation}   
W_{j,i}=\tilde{Z}_{j,i}- \frac{1}{k-1} \sum_{u=k+1 }^{2k-1}\tilde{Z}_{j,u}
\end{equation}
for $i=2, \cdots , k$. All these statistics have have the same distribution under $H_0$. 
%Here, $W_{j,1}$ large indicates significance and $W_{j,i}$ large for $i>1$ are only noise. The motivation for
%introducing multiple set of knockoffs is to reduce the noise in the knockoffs. 
For $k=2$ this formula gives the expressions  $ W_{j,1}= Z_{j,1}- \tilde{Z}_{j,3}$ and $ W_{j,2}= \tilde{Z}_{j,2}- \tilde{Z}_{j,3}$
which is slightly different than in
\cite{Candes17}. Here, we have chosen $k-1$ elements in the sums (2) and (3). We could have chosen any other integer number.
For $k>2$,  $W_{j,i}$ is not symmetric in contrast to Candes et al.  \cite{Candes17}.
A large value of  $W_{j,1}$ indicates that the variable $j$ is significant, i.e. $\beta_j \ne 0$ while
  a large value of  $W_{j,i}$ for $i>1$ is only due   to randomness.  We consider
 $W_{j,1}>T$ for a threshold $T>0$ as an indication that component $j$ is significant.  Hence,  we must count the number of
$W_{j,i}>t>0$ for different thresholds $t$  for $i=1$ and $i>1$ respectively. In order to control the
 false discovery rate, FDR, it is essential that $W_{j,1}$ and  $W_{j,i}$ have  the same density for $i=2,3,\cdots , k$
 for each $j=1,2,\cdots , p $ when $j \in H_0$.   Inspired by Barber and Candes \cite{Barber15} and 
 Candes et al.  \cite{Candes17}  we define the threshold
%two different methods for classification of the hypotheses. The standard knockoff alternative is to define
\begin{equation}
T = min \{ t>0: \frac{   \# \{ j,i>1: W_{j,i} \geq t \} }{  \# \{ j: W_{j,1} \geq t \} (k-1) } \le q \}  
\end{equation}

We define $T= \infty  $ if the set described above are empty. 
We reject that  $j \in H_0$ if  $W_{j,1} \geq T$.

Define $I(W_{j,i}\geq T)=1$ if $W_{j,i} \geq T$ and $0$ otherwise. Then we may formulate the following Theorem bounding the false
discovery rate FDR.

\begin{theorem}
  Assume
  \begin{equation}   
   E \frac{ I( W_{u,1} \geq T ) }{ (\sum_{j=1}^{p} I( W_{j,1} \geq T )) \vee 1  }  \leq E   \frac{ I( W_{u,i} \geq T ) }{
     ( \sum_{j=1}^{p} I( W_{j,1} \geq T )) \vee 1  } 
  \end{equation}
  for $i=2,3,\cdots ,k$ and $u=1,2, \cdots, p$ assuming $ u \in H_0$.
Then  
$$ FDR= E \frac{ \sum_{j=1, j \in H_0}^{p} I( W_{j,1} \geq T ) }{(\sum_{j=1}^{p} I( W_{j,1} \geq T ))  \vee 1 } \leq q $$
  for all integers $k>1$. The expectation is taken over the noise in the response $\varepsilon $ and the knockoffs $\tilde{X}$ while
  keeping the covariates $X$ fixed. 
\end{theorem}
\vspace{5mm}

It is necessary  to verify the assumption (6). If $W_{j,i}$ have the same properties for all values of $j$ and $i$, the assumption is
satisfied. This follows from the following calculation. We may assume
$  \sum_{j=1}^{p} I( W_{j,1} \geq t )) \geq 1$. If this is not the case, the left hand side is vanishing and the assumption is
satisfied trivially. When we take the sum over all possible left hand sides of inequality (6), we get:
$$  \sum_{v=2}^{k} \sum_{s=2}^{k} \sum_{u=1}^{p}\frac{ I( W_{u,1} \geq t ) }{ \sum_{j=1}^{p} I( W_{j,1} \geq t )  } =
 (k-1)^{2}  \frac{ \sum_{u=1}^{p} I( W_{u,1} \geq t ) }{ \sum_{j=1}^{p} I( W_{j,1} \geq t )  }= (k-1)^{2} .  $$
When we take the sum over all possible right hand sides, we get: 
%$$  \sum_{v=1}^{k} \sum_{s=1}^{k} \sum_{u=1}^{p}\frac{ I( W_{u,s} \geq t ) }{ \sum_{j=1}^{p} I( W_{j,v} \geq t )  } \geq k^{2} .  $$
%This is the sum over the $p k^{2}$ possible right hand sides of (6) while the sum over the $p$ possible right hand sides for (6) is $1$.
%This inequality follows from
$$  \sum_{v=2}^{k} \sum_{s=2}^{k} \sum_{u=1}^{p}\frac{ I( W_{u,s} \geq t ) }{ \sum_{j=1}^{p} I( W_{j,v} \geq t )  } =
\sum_{v=2}^{k} \sum_{s=2}^{k} \frac{ \sum_{u=1}^{p} I( W_{u,s} \geq t ) }{ \sum_{j=1}^{p} I( W_{j,v} \geq t )  } \geq  $$
$$ (k-1) \sum_{v=2}^{k}  \frac{ \sum_{u=1}^{p} I( W_{u,v} \geq t ) }{ \sum_{j=1}^{p} I( W_{j,v} \geq t )  } =   (k-1)^{2} .  $$
When we set $ a_v =  \sum_{u=1}^{p} I( W_{u,v} \geq t )$,  the inequality follows from
$$ \sum_{v=2}^{k} \frac{ a_{p(v)}  }{ a_{v} } \geq \sum_{v=2}^{k}  \frac{  a_{v}  }{ a_{v}} =k-1   $$
   for any set of $a_v >0 $ for $v=2,3 \cdots, k$ and any permutation vector $p(v)$.

 If $W_{j,i}$ do not have the same properties for all values of $j$ and $i$, assumption (6) depends on how the test statistics $Z_{j,i}$ are defined. 
The variables $W_{j,i}$ have the same distribution for $i=1,2,\cdots , k$ under $H_0$ and hence,
the left and right hand side of (6) are quite similar.
All the test statistics $W_{j,i}$ are correlated. It is easy to estimate the assumption  numerically by calculating
$$  \frac{ I( W_{u,s} \geq t ) }{ (\sum_{j=1}^{p} I( W_{j,v} \geq t )) \vee 1  }   $$
for different values of $t>0$ where   $s=2,3 \cdots, k$ and  $v=2,\cdots, k$ and test out whether  $s=v$ on average gives smaller
values than for $s \ne v$. When we have $s>1$ and $v>1$, it is easy to estimate the inequality  under the $H_0$ assumption.

Another  argument for the assumption (6) is
that $I( W_{j,1} \geq T)$ is part of the denominator and implying that when the nominator is positive, the
 denominator has at least one positive term increasing the expected value of the denominator  making the fraction smaller. For many
test statistics $Z_{j,i}$, it is more likely with a positive correlation between  $I(W_{j,1} \geq T)$ for $j=1,2,\cdots, p$
also making the assumption (6) more likely.

%The same argument may also be applied to the test statistics in the panning for gold article. The gives slightly stronger FDR bound than proved
%in the article, but this depends on that Property (3) is satisfied.
\vspace{10mm}

{\bf Proof }

We may write the definition of $T$ as
$$T = min  \{ t>0: \frac{ K_t }{  D_t } \leq q \}$$  
where $K_t = (\sum_{j=1, i=2}^{p,k} I( W_{j,i} \geq T ))/(k-1)$ and $D_t=\sum_{j=1}^{p} I( W_{j,1} \geq T ) $.
Note that $T$ is defined as a function based on the
 variables  $W_{j,i}$ which again depend on the noise in the response $\varepsilon $, the knockoffs $\tilde{X}$ and  the covariates $X$.
%We consider the covariates $X$ as fixed and denote $d=(\varepsilon, \tilde{X})$.
If the inequality $\frac{ K_t }{  D_t } \leq q$ is
not satisfied for any value of $t>0$, then  $T= \infty  $ implying that the Theorem is satisfied trivially. 

Define $R_t= \sum_{j=1, j \in H_0}^{p} I( W_{j,1} \geq t ) $.   %\# \{ j \in H_0: W_{j,0} \geq t \}$. 
Then we have 

$$ FDR = E\frac{R_T}{D_T \vee 1} =  E \sum_{j=1, j \in H_{0} }^{p} \frac{I(W_{j,1}\geq T)}{D_T \vee 1} \leq
E \sum_{i=2,j=1, j \in H_{0} }^{k,p} \frac {I(W_{j,i}\geq T)}{(D_T \vee 1) (k-1)} \leq $$
$$ E \sum_{i=2,j=1 }^{k,p} \frac {I(W_{j,i}\geq T)}{(D_T \vee 1)(k-1)} =   E\frac{K_T}{D_T \vee 1}
\leq  E(\frac{ q D_T }{ D_T \vee 1})  \leq  q.   $$
We first split $R_T$ in the separate terms and use assumption (6).
Later we utilize that the definition of $T$ implies that $K_T \leq q D_T$.
This  proves the Theorem.

\section{Test 1}

%\begin{figure}               
%\begin{center}             
%\includegraphics[width=0.5\linewidth]{FDR2N.pdf}   
%\caption{ Test 1: False discovery rate as function of number of knockoffs, $k$ for the three alternative methods described. 
%  $W$ is continuous line, $V$ is dashed line and $Q$ is dotted line.  }   
%\label{FDR}
%\end{center}             
%\end{figure}

\begin{figure}               
  \begin{center}
    \includegraphics[width=0.5\linewidth]{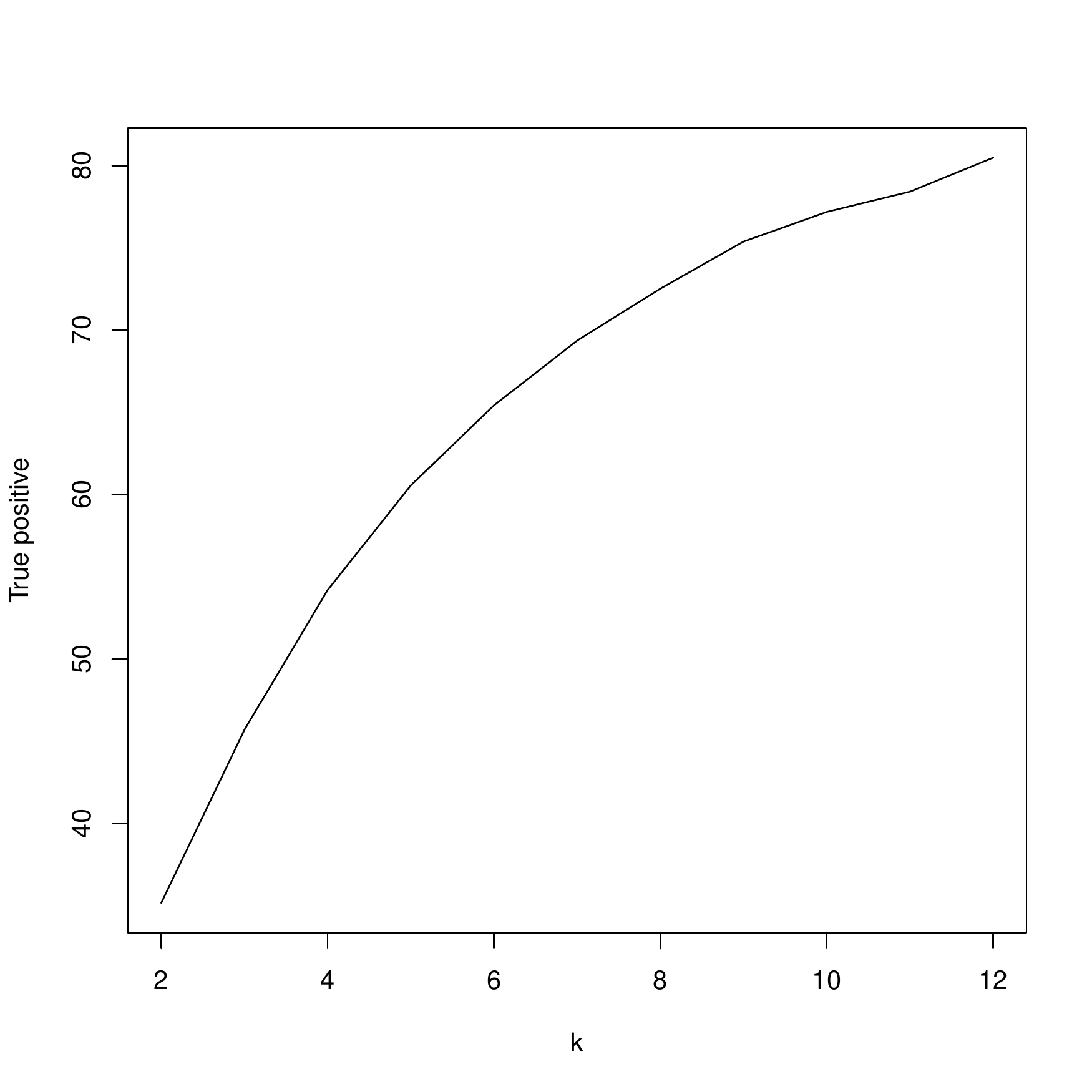}   
\caption{ Test 1: Number of true positive as a function of number of test statistics $k$.  }   
\label{fig-truepos}
\end{center}             
 \end{figure}

The following example illustrates the effect of multiple knockoffs. Let $Z_{j,1} = X_j \sim N(\mu_j,1)$ for 
$j=1, \cdots , p$ be the data  where $\mu_j=a>0$ for $ j \notin  H_0 $ and otherwise $\mu_j=0$. The other  $Z_{j,i} = X_j \sim N(0,1)$
are noise for $i=2,3,\cdots , 2k-1$.
The problem is to identify the components $j$ with $\mu_j=a>0$. In the simulation we let $p=5 000$, $a=2$, $p/10$
components in $H_0$. We also let the false discovery rate be 10 \%.  Simulation shows that FDR is slightly below 10\% as is should.
%Figure \ref{FDR} shows the false discovery rate, i.e. the average false discovery rate for all the simulations
%for the three different methods.
Figure \ref{fig-truepos} shows the number of true positive. Notice that it increases significantly from  $k=2$,
We also verify assumption (6) numerically in this example. The right hand side is slightly larger than
the left hand side.

 \section{acknowledgements}
   The authors thanks Matteo Sesia for valuable comments. 

%--------------- bibliography                   


\begin{thebibliography}{10}                   
  
\bibitem{Barber15}% and Tasche, 2002}%{Acerbi02} %[Acerbi and Tasche, 2002]  
Barber, R. F. and Candes, E.J., Controlling the false discovery rate via Knockoffs, Annals of Statistics
43 (5) 2055-2085, 2015.

\bibitem{Barber17}% and Tasche, 2002}%{Acerbi02} %[Acerbi and Tasche, 2002]  
Barber, R. F. and Candes, E.J., A knockoff filter for high-dimensiaonal selective inference
, arXiv 2017.

\bibitem{Barber18}% and Tasche, 2002}%{Acerbi02} %[Acerbi and Tasche, 2002]  
Barber, R. F., Candes, E.J. and Samworth, R., Robust Inference with Knockoffs, arXiv 2018.


\bibitem{Candes17}% and Tasche, 2002}%{Acerbi02} %[Acerbi and Tasche, 2002]  
Candes, E., Fan, Y., Janson, L. and  Lv, J., Panning for Gold: Model-free Knockoffs for High-dimensional 
Controlled Variable Selection,  arXiv 2017
  


\begin{comment}  
\bibitem{Artzner97}%[Artzner et al, 1997]  
Artzner, P., Delbaen, F., Eber, J.-M. Heath, D., Thinking coherently, Risk 10 (11) 1997.  
  

\bibitem{Tasche07}%[Tasche, 2007]   
Tasche, D. Euler allocation: Theory and practice, Working paper,   2007.  
  
\bibitem{Urban04}%[Urban et al., 2004]   
Urban, M, Dittrich, J., Kluppelberg, C., and Stolting, R., Allocation of risk capital to insurance  
portfolios, Blatter der DGVFM, 26, 389-406, 2004  
  
\end{comment}  
                   
%\bibitem{Atkinson97} P. Embrechts, A. J. McNeil and D. Straumann, Correlation: PitFalls and alternatives, Risk 12:69-71, 1999.  
  
\end{thebibliography}
\end{document}